\documentclass[allclo]{FBSart}
\usepackage{epsfig}
\newcommand{\beq}{\begin{eqnarray}}
\newcommand{\eeq}{\end{eqnarray}}
\title{The large $N$ limit from the lattice\footnote{Presented at Light-Cone 2004, Amsterdam, 16 - 20 August}}
\author{Biagio Lucini}
\institute{Institute for Theoretical Physics, ETH Z\"urich, CH-8093 Z\"urich, Switzerland}
\runningauthor{Biagio\, Lucini}
\runningtitle{LC 2004}
\begin{document}
\maketitle
\begin{abstract}
A numerical study of the string tension and of the masses
of the lowest-lying glueballs is performed in SU($N$) gauge theories
for $2 \le N \le 8$ in D=3+1 and $2 \le N \le 6$ in D=2+1.
It is shown that for the string tension a smooth $N \to \infty$
limit exists that depends
only on the 't Hooft coupling $\lambda = g^2 N$. An extrapolation of the masses
of the lightest glueballs to $N = \infty$ using a power series in $1/N^2$
shows that the leading correction to the infinite $N$ value accounts
for finite $N$ effects for $N$ at least as small as 3 and all the
way down to $N=2$ in many cases. $k$-string tension ratios  and possible
issues connected with correlation functions at large $N$ are also discussed.
\end{abstract}
\section{Introduction}
A better understanding of the large $N$ limit of SU($N$) gauge theories 
is an important prerequisite for advances in several branches of 
particle physics. It has been shown that accurate numerical results
for observables in SU($\infty$) can be attained by extrapolating to
$N = \infty$ values obtained at finite $N$~\cite{mt-21,blmt-glue,pisa-Q}.
The extrapolation is based on the diagrammatic expectation that observables
at finite $N$ differ from the corresponding observables at $N=\infty$ 
by corrections that are described by a power series in
$1/N^2$~\cite{largeN}.
In this work we will present results for the string tension, $k$-string
tension ratios and glueball masses in 3+1~\cite{blmtuw-glue}
and 2+1 dimensions~\cite{blmt-21}.
\begin{figure}[t]
\begin{center}
\label{fig1}
\epsfig{figure=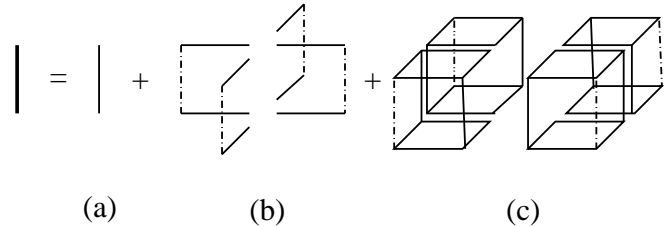,height=3cm}
\caption{In our scheme, a smeared link is obtained by adding to the original
link (a) the parallel transported nearest neighbour links (b) weighted
by a coefficient $\alpha$ and the next to nearest neighbour links parallel
transported along all possible paths on the elementary cube (c) weighted
by another coefficient $\beta$. For a judicious choice of $\alpha$ and $\beta$
it is possible to build operators with a 98\% overlap on the physical
states.}
\end{center}
\end{figure}
\section{The method}
We have studied numerically on the lattice SU($N$) gauge theories for
$2 \le N \le 8$ in 3+1 dimensions and $2 \le N \le 6$ in 2+1 dimensions.
For any given $N$ we have used the Wilson action
\beq
S = \beta \sum _{i, \mu, \nu} \left( 1 - \frac{1}{2N} \mbox{Tr}
\left( U_{\mu \nu}(i) + U_{\mu \nu}^{\dag}(i) \right) \right) \ ,
\eeq
where the $U_{\mu \nu}(i)$ are path ordered products of SU($N$) link
variables along the elementary lattice squares ({\em plaquettes}) and 
$\beta = 2 N/(g^2 a^{4-D})$, with $g$ the coupling of the theory,
$D$ the dimensionality of the system and $a$ the lattice spacing.
The values of $\beta$ used in our study have been chosen above
the bulk phase transition point~\cite{blmt-glue}.

Masses can be extracted from the exponential decay of correlation functions
built with traces of path ordered products of functions of link
variables. In order to have a significant overlap with physical states,
these functions must be smooth on the
characteristic distance scale of those states. Traditionally, iterated
smearing~\cite{smear} and blocking~\cite{block} are used to achieve this
result. In our work in D=3+1 we have
used a combination of improved smearing (see Fig.~\ref{fig1}) and
blocking, while simple blocking has been used in D=2+1.
For each state of interest, a set of operators carrying the required
quantum numbers has been built out of smeared and blocked variables
and the corresponding correlation matrix has been computed. Masses have been
obtained by a variational calculation performed on this matrix. More details
on the numerical methods we have used can be found
in~\cite{blmtuw-glue,blmt-21}.

\section{Correlators and large $N$}
There is a potential problem with the application of the method outlined
in the previous section at large $N$. In fact, from general arguments,
as $N \to \infty$ we expect factorisation of correlators, which implies that
the connected part of the 2-point function (from which we extract masses) is
${\cal O}(1/N^2)$.
At large enough $N$ then this signal might become invisible.
However, what determines the accuracy of the calculation is not the
absolute magnitude of the signal, but the ratio signal over noise.
Large $N$ arguments~\cite{blmtuw-glue} show that this ratio
is independent of $N$, and this expectation is confirmed by our
data. Thus, despite the fact that the signal disappears as $1/N^2$,
we are able to compute correlators with an accuracy that is independent
of $N$.
\begin{figure}[ht]
\begin{center}
\epsfig{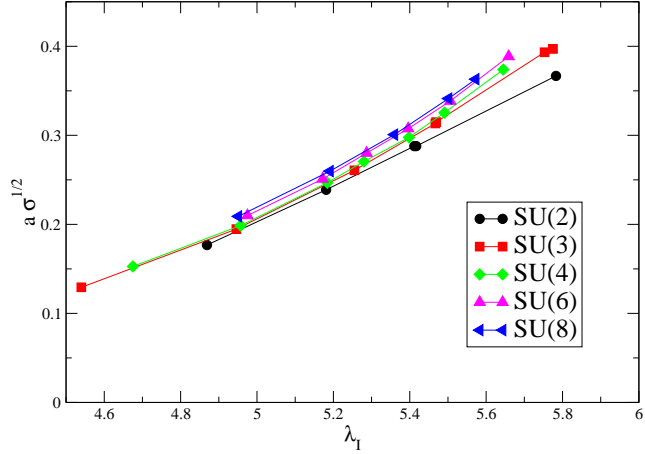}
\caption{$a \sqrt{\sigma}$ as a function of $\lambda_I$ at various $N$
in D=3+1.}
\label{fig2}
\end{center}
\end{figure}
\begin{figure}[t]
\begin{center}
~\\~\\
\epsfig{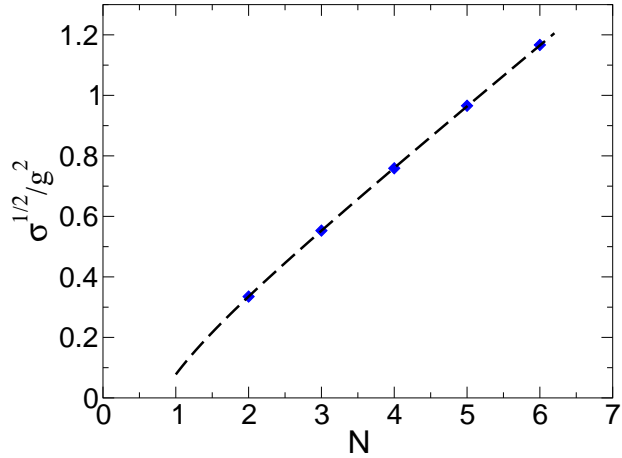}
\caption{The square root of the string tension expressed in units of $g^2$
as a function of $N$ in D=2+1. The dashed line is our best fit
to the data.}
\label{fig3}
\end{center}
\end{figure}
\section{String tension}
The string tension
$\sigma$ can be obtained from torelon correlation functions~\cite{blmt-glue}.
Diagrammatic arguments suggest that equivalent physics across SU($N$)
gauge groups is obtained at fixed $\lambda = g^2 N$. Since $\lambda$
depends on the scale at which the physics is probed, equivalent $\lambda$
are expected at the same physical scale, or equivalently the same scale
is identified by a given $\lambda$ at any large enough value of $N$.
Although this is not an ideal choice, on the lattice we can easily
verify this statement on length scales of the order of $a$.
At this scale lattice artifacts resulting from a naive definition of the
lattice coupling can spoil the comparison.
To limit the effects of those artifacts
we have used the mean field improved coupling $\beta_I = \beta/<U_{\mu\nu}>$,
where $<U_{\mu\nu}>$ is the average plaquette. Fig.~\ref{fig2} shows a
plot of $a \sqrt{\sigma}$ as a function of $\lambda_I = 2 N^2/\beta_I$
in D=3+1. This plot suggests that a smooth large $N$ limit
for $\lambda_I$ exists at all values of $a \sqrt{\sigma}$.

In D=2+1 the lattice spacing $a$ can be traded with the coupling
$g^2$, which now has dimensions of mass. Our data are displayed in
Fig.~\ref{fig3}. From diagrammatic arguments, we expect that
up to ${\cal O}(1/N^2)$ corrections, in the continuum
limit $\sqrt{\sigma}/g^2 \propto N$. A fit gives
\beq
\frac{\sqrt{\sigma}}{g^2 N} = 0.19755(35) - \frac{0.120(3)}{N^2} \ .
\eeq
\begin{figure}[ht]
\begin{center}
\label{fig4}
\vspace{0.7cm}
\epsfig{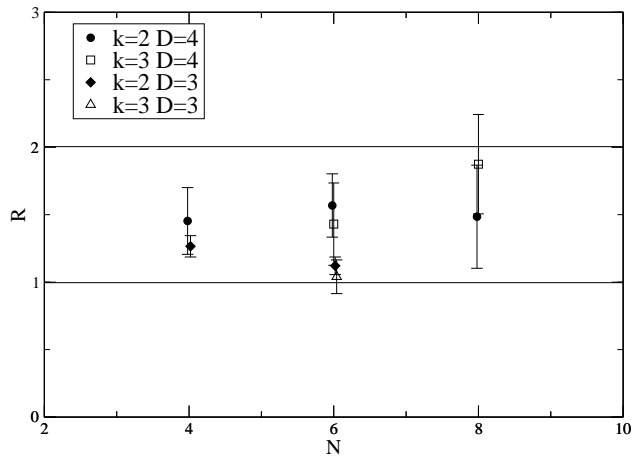}
\caption{Ratios of the tensions of strings of various ${\cal N}$-alities over
the tension of the fundamental string. The scale on the $y$ axis has been
chosen in such a way that the values $R=1$ and $R=2$ always correspond
respectively to Casimir scaling and to the sine formula. Those values are
indicated by the horizontal lines in the plot.}
\end{center}
\end{figure}
\section{$k$-strings}
Strings connecting sources in a representation of rank (or
{\em ${\cal N}$-ality}) $k < N$ of SU($N$) with their antisources
are known in literature as {\em $k$-strings}. Since strings associated to
sources of different ${\cal N}$-ality can not mix, at fixed $N$ there
are int[$N/2$] stable $k$-strings. The values of their tensions are
determined by the underlying dynamics of colour confinement.

$k$-string tensions can be extracted from correlators of multi-wrapping
torelons and powers of them that under the centre transformation
$e^{2 \pi i/N}$ acquire the phase $e^{2 \pi i k/N}$. Our numerical
calculations show that the ratios $R$ of tensions of strings connecting
sources of ${\cal N}$-ality $k$ and $N-k$ over the tension
of the fundamental string lay between the Casimir scaling ($R=k(N-k)/(N-1)$)
and the sine formula $(R=\sin(k \pi/N)/\sin(\pi/N))$ predictions in
D=3+1~\cite{blmtuw-glue,blmt-kstring} (see also~\cite{PisaK}, where a
better agreement with the sine formula is claimed), while in D=2+1 Casimir
scaling seems to be favoured~\cite{blmt-kstring}.
Although the corresponding numerical values are very close,
those formulae hide very different scenarios: while the sine formula can be
expanded in a series that contains only even powers of $1/N$, the first term
in the expansion of the Casimir prediction is proportional to $1/N$, and this
is hardly accommodated by diagrammatic arguments~\cite{mt,aams}.
Even if both formulae at small $N$ are expected to receive sizeable
corrections that undermine the numerical comparison, the question of the
power dependence of the leading correction in a $1/N$ expansion still stands.
Current numerical data are not accurate enough to provide a reliable answer
to this problem.
\begin{table}
\label{table1}
\begin{center}
\begin{tabular}{|l||c|c||c|c|}\hline
~ & \multicolumn{2}{|c|}{D=2+1} & \multicolumn{2}{|c|}{D=3+1} \\
\hline \hline
state & $d_0$ & $d_1$ & $d_0$ & $d_1$ \\ \hline \hline
$0^{++}$         & 0.8116(36) &  -0.090(28) & 3.28(8)  & 2.1(1.1) \\
$0^{++\ast}$     & 1.227(9)   &  -0.343(82) & 5.93(17) & 2.7(2.0) \\
$0^{++\ast\ast}$ & 1.65(4)    &  -2.2(7)    & --       & --       \\
$0^{--}$         & 1.176(14)  &  0.14(20)   & --       & --       \\
$0^{--\ast}$     & 1.535(28)  &  -0.35(35)  & --       & --       \\
$2^{++}$         & 1.359(12)  &  -0.22(8)   & 4.78(14) & 0.3(1.7) \\
$2^{++\ast}$     & 1.822(62)  & -3.9(1.3)   & --       & --       \\
$2^{--}$         & 1.615(33)  & -0.10(42)   & --       & --       \\
\hline
\end{tabular}
\end{center}
\caption{Fit parameters for the extrapolation to the $N=\infty$ limit of
masses of glueballs in D=2+1 according to formula~(\ref{massfit21})
and in D=3+1 according to formula~(\ref{massfit31}). The quantum numbers
of the states are indicated with the standard notation $J^{PC}$ and higher
excitations in a given channel are denoted with a corresponding
number of $\ast$.}
\end{table}
\section{Glueball masses}
Glueball masses are extracted from correlators of products of smeared and
blocked links along contractible paths. The transformation properties of
the paths under rotation, parity and charge conjugation determine the quantum
numbers of the states. We have fitted our data with the ansatz
\beq
\label{massfit21}
m_N / g^2 N = d_0 + d_1/N^2
\eeq
in D=2+1 and
\beq
\label{massfit31}
m_N / a \sqrt{\sigma} = d_0 + d_1/N^2
\eeq
in D=3+1. The results of our fits are reported in Table 1.
In D=2+1, the ansatz works for
all the states all the way down to $N=2$. In D=3+1 we draw
similar conclusions for the $0^{++}$ glueball, while a correction ${\cal
O}(1/N^2)$ turns out not to be enough to describe the SU(2) numerical data for
the $0^{++\ast}$ and the $2^{++}$ states. Note that while both in 2+1 and
3+1 dimensions we can extract accurate values for the adimensional masses
in the $N \to \infty$ limit, the values of the coefficients of the leading
correction is determined unambiguously only in the D=2+1 case. In D=3+1
we can however estimate their order of magnitude. Our results support the
conclusion that at finite $N$ the deviation from the $N \to \infty$ value
is a genuine ${\cal O}(1/N^2)$ correction.
\section{Conclusions}
Our investigation shows that a smooth large $N$ limit exists for several
physical observables and the physics at finite $N$ is correctly described
by the leading expected correction all the way down to at least $N=3$.
In particular, we have discussed our results for the masses of the lightest
glueballs and for the string tension in both D=2+1 and D=3+1. Unlike
early works on this subject, in most cases we were able to provide accurate
and reliable numerical estimates for the $N = \infty$ value and the leading
correction. An interesting problem is whether for some observables the approach
to the $N = \infty$ limit can be described by a $1/N$ correction. On the light
of present results, this possibility can not be ruled out for ratios
of $k$-string tensions.
\section*{Acknowledgement}
This work was done in collaboration with M. Teper and U. Wenger.
The author would like to thank B. Bakker and S. Dalley
for the invitation to speak at the conference and P. van Baal, S. Brodsky,
S. Dalley and M. Karliner for discussions.
\end{document}